\documentclass[english]{iopart}
				   
\begin{document}

\title[Painlev{\'{e}} versus Fuchs]
{\Large
Painlev\'e versus Fuchs\footnote{Dedicated to the centenary of the publication of the
Painlev\'e VI equation in the Comptes Rendus de l'Acad\'emie des Sciences de Paris
by Richard Fuchs in 1905.}}

\author{S. Boukraa$^\dag$,  S. Hassani$^\S$,
J.-M. Maillard$^\ddag$, B. M. McCoy$^\P$,
J.-A. Weil$^\star$ and N. Zenine$^\S$}
\address{\dag Universit\'e de Blida, D\'epartement d'A{\'e}ronautique,
 Blida, Algeria}
\address{\S  Centre de Recherche Nucl\'eaire d'Alger, \\
2 Bd. Frantz Fanon, BP 399, 16000 Alger, Algeria}
\address{\ddag\ LPTMC, Universit\'e de Paris 6, Tour 24,
 4\`eme \'etage, case 121, \\
 4 Place Jussieu, 75252 Paris Cedex 05, France} 
\address{\P Institute for Theoretical Physics,
State University of New York,
Stony Brook, USA}
\address{$\star$ XLIM, Universit\'e de Limoges,
123 avenue Albert Thomas,
87060 Limoges Cedex, France} 
\ead{maillard@lptmc.jussieu.fr, maillard@lptl.jussieu.fr, 
sboukraa@wissal.dz, njzenine@yahoo.com}

\begin{abstract}
 
The sigma form of the Painlev{\'e} VI equation contains 
four arbitrary parameters and
generically the solutions can be said to be genuinely ``nonlinear''
because they do not satisfy linear differential equations of finite
order. However, when there are certain restrictions on the four parameters
there exist one parameter families of solutions  which do satisfy
(Fuchsian) differential equations of finite order. We here study this
phenomena of Fuchsian solutions to the Painlev{\'e} equation with a
 focus on the particular PVI equation which is satisfied by
the diagonal correlation function $C(N,N)$ of the  Ising model. We
obtain Fuchsian equations of order $N+1$ for $C(N,N)$  and
show that the equation for $C(N,N)$ is equivalent  to the 
$N^{th}$ symmetric power of the equation for the elliptic integral $E$.
 We show
that these Fuchsian equations correspond to rational algebraic curves
with an additional Riccati structure and we show that the Malmquist 
Hamiltonian $p,q$ variables are rational functions in complete elliptic
integrals. Fuchsian equations for off diagonal correlations
$C(N,M)$ are given which extend our considerations to discrete
generalizations of Painlev{\'e}.

\end{abstract}
\vskip .3cm

\noindent {\bf PACS}: 02.30.Hq, 02.30.Gp, 02.30.-f, , 02.40.Re, 05.50.+q, 05.10.-a, 04.20.Jb
\vskip .2cm
\noindent {\bf AMS Classification scheme numbers}: 33E17, 33E05, 33Cxx,  33Dxx, 14Exx, 14Hxx, 34M55, 47E05,
 34Lxx, 34Mxx, 14Kxx
\vskip .3cm
 {\bf Key-words}:  sigma form of Painlev\'e VI, two-point 
correlation functions of the Ising model,  Fuchsian 
linear differential equations,  
 apparent singularities, rigid local systems, differential
 Galois group, B\"acklund transformations, complete elliptic 
integrals, rational curves, Riccati equations

\section{Introduction}

\label{intro}

The correlation functions of the Ising model were first calculated by
Kaufman and Onsager \cite{kau-ons-49}  
in terms of determinants whose elements are
certain hypergeometric functions. For this reason it follows from a
theorem on holonomic functions \cite{holonomic} that they must satisfy
linear ordinary differential equations. However, these correlations
also have a remarkable connection with nonlinear equations as well.
The first such result
was the expression as $T \rightarrow T_c$ of the scaled correlation 
function in 
terms of a PIII function by Wu, McCoy, Tracy and Barouch
\cite{wu-mc-tr-ba-76} in
1976. Subsequently in 1980 it was shown for arbitrary fixed $T$ 
by Jimbo and Miwa
\cite{jim-miw-80} 
that the
diagonal correlation $C(N,N)$ is given in terms of a PVI function and 
by McCoy, Wu  \cite{mcc-wu-80} and Perk \cite{perk-80} that the 
correlation at a general position  $C(M,N)$
and its ``dual'' $C^{*}(M,N)$ satisfy some remarkable quadratic identities,
or double recursions 
which  are discrete generalizations of the Painlev\'e ODE's. 

The Painlev{\'e} representation of the correlation functions is by now
well known but, curiously enough, almost nothing is known about the
corresponding linear equations beyond the fact that the diagonal
correlation function $C(1,1)$ is a particular case of the
hypergeometric function. In this paper we will study these linear
equations for the Ising correlation functions and the much more
general question of when solutions of the PVI equation will satisfy
Fuchsian differential equations.

The most  general four parameter dependent sigma
form of Painlev\'e VI can be written as \cite{okamoto-79, Forrest}
\begin{eqnarray}
\label{pvi}
&&\zeta'(t(t-1)\zeta'')^2\, +(2\zeta'(t\zeta'-\zeta)\, -\zeta'^2
  -v_1v_2v_3v_4)^2\nonumber\\
&&=\, (\zeta'+v_1^2)
(\zeta'+v_2^2)\, (\zeta'+v_3^2) \, (\zeta'+v_4^2)   \qquad \hbox{with:}
\end{eqnarray}
\begin{equation}
\zeta=t(t-1){d \ln \tau\over dt}\, +K_1\, t\, +K_2 \qquad \hbox{where:}
\end{equation}
\begin{equation}
K_1\,= \,\, v_1v_2-v_1v_3-v_2v_3, \qquad \qquad \hbox{and:}
\end{equation}
\begin{equation}
K_2\, =\,\,  -{1\over 2}(v_1v_2-v_1v_3-v_1v_4-v_2v_3-v_2v_4+v_3v_4)
\end{equation}
This is a second order nonlinear equation which allows
branchpoints only at the three points $t=0, 1, \, \infty$ and
locally near these singularities the function $\tau$ has, following Jimbo's expansions~\cite{jim-82},
 an expansion of the form
\begin{equation}
\tau=x^{p_j}\sum_{k=-\infty}^{\infty} \, x^{k^2+k\alpha}\sum_{n=0}^{\infty}
\eta^{-k} \cdot  a_j(n,k;\alpha)\cdot x^n
\label{expand}
\end{equation}
where $x$ is the local variable at $t=0,1,\infty,$  
and two boundary conditions for the second order PVI equation 
specified by $\alpha$ and  $\eta$ will in general be different at the
three singularities. 
The coefficients $a_j(n,k;\alpha)$ depend on the value of
$j=0,1,\infty$ and satisfy
$a_j(n,-k,\alpha)=a_j(n,k,-\alpha)$
and we note that
\begin{eqnarray}
p_0=&&\{\alpha^2-(v_1+v_2-v_3-v_4)^2\}/4\\
p_1=&&\{\alpha^2-(v_1+v_2-v_3+v_4)^2\}/4 \\
p_{\infty}=&& \alpha^2/4+ K_1
\end{eqnarray}
Comparison of (\ref{expand}) with the well known expansion of Jimbo
\cite{jim-82} reveals that many of the coefficients in Jimbo's expansion
vanish identically. Several $a_j(n,k,\alpha)$ are explicitly given in
Sec. 2.1.

In general the local expansion (\ref{expand}) has an infinite number of
confluent singularities which indicates that it  cannot satisfy
a {\em linear} differential equation.
 Therefore even though the most general solution of the PVI
 equation cannot satisfy
 a {\em linear} equation, the  specific boundary conditions which specify the
 solution to be the physical diagonal correlation function of the
 Ising model
 will allow a Fuchsian equation of 
{\em order generically greater than two} to be satisfied.

In this paper we study this phenomena of the 
existence of boundary conditions
for which solutions of certain PVI equations satisfy Fuchsian differential 
equations\footnote{For a warm-up on  Painlev\'e VI, 
sigma form of Painlev\'e VI, 
 and on the question of the holonomic solutions
inside Painlev\'e VI we recommend two magnificent papers in French, 
one by Garnier~\cite{Garnier}  
and the other one by Okamoto~\cite{okamoto-79} (see also in English~\cite{okamoto-87}).}. 
There are several ways in which this phenomenon may occur. One way is
that conditions can be found on the four parameters $v_k$ and on
$\alpha$ such that the general local expansions at $t=0, 1, \infty$  
degenerate by having 
the coefficients $a_j(n,k;\alpha)$
all vanish if $k$ is sufficiently large. This will give a one
parameter family of solutions which has only a finite number of
confluent singularities. 
We study this mechanism in
detail in Sec. 2.1. However, there may also exist one parameter
families which cannot be obtained from the two parameter families
(\ref{expand}) by specialization. An example of this is given in Sec. 2.4.

For concreteness we will consider in detail the specific PVI equation for 
the diagonal Ising correlation obtained
by Jimbo and Miwa \cite{jim-miw-80}:
\begin{eqnarray}
\label{jimbo-miwa}
&&\Bigl( t\, (t-1) \sigma^{''} \Bigr)^2\, =\,\,   \\
&&\quad N^2 \cdot \Bigl( (t-1) \sigma^{'} -\sigma \Bigr)^2\,\,  -4\,\sigma^{'}
\Bigl((t-1)\sigma^{'}-\sigma -1/4   \Bigr)\Bigl(t\sigma^{'}-\sigma   \Bigr) \nonumber 
\end{eqnarray}
which is obtained from (\ref{pvi}) by setting
\begin{eqnarray}
\label{isingv}
&& v_1\, =\, v_4\,= \, N/2,\quad  v_2\,= \, \, (1-N)/2,\quad v_3\, =\, (1+N)/2 \\
&& \sigma\, =\, \zeta\, +N^2t/4-1/8
\end{eqnarray}
The diagonal  $C_N=C(N,N)$ is related to  $\sigma$ for $T >T_c$ by
\begin{eqnarray}
\label{sigma-bas}
&&\sigma(t)\,=\,\, t(t-1) \cdot {\frac{d}{dt}}\log(C_N)\,
-1/4\nonumber\\
&&{\rm with}~~t=\, \, \Bigl( \sinh(2J_v/kT)\cdot \sinh(2J_h/kT) \Bigr)^2 <1
\end{eqnarray}
where $\, C_N\, \simeq \, t^{N/2}$ when $\, t \, \rightarrow \, 0$, 
and for $T <T_c$ by
\begin{eqnarray}
\label{sigma-haut}
&&\sigma(t)\,=\,\, t(t-1)  \cdot{\frac{d}{dt}}\log(C_N)\,
-t/4\nonumber\\
&&{\rm with}~~t=\, \, \Bigl( \sinh(2J_v/kT)\cdot \sinh(2J_h/kT) \Bigr)^{-2} <1
\end{eqnarray}
where $\, C_N\, \simeq \, 1$ when $\, t \, \rightarrow \, 0$,
and where the  variable  
$J_v$ ($J_h$) is the Ising model vertical (horizontal) coupling constant.
The detailed specification of the behavior of $\, \sigma(t)\,$ near $\, t=0$
needed to uniquely specify  $\, C_N\,$ as the diagonal Ising 
correlation function are sketched in section (\ref{nontriv}) (see for instance equation (\ref{boundary})).
Do note that since all the calculations of this paper are systematically
checked with high-temperature expansions when available, we introduce 
a variable $\, t$ which is the inverse  of the one of Jimbo and Miwa \cite{jim-miw-80}.
For integer $N$ the equation (\ref{jimbo-miwa}) is in the class of 
so called ``classical'' equations \cite{Forrest} 
which are known to generate Toeplitz
determinants whose elements are hypergeometric
 functions~\cite{okamoto-79,Forrest,okamoto-87}. 
We present the Fuchsian equations satisfied by $C_N$ for small
values of $N$ in Sec. 2. These equations have remarkable structure and 
in Sec. 2.2 we show that the associated $\, N+1$ order differential
operators are homomorphic to the  $N$-th symmetric power
of the second order differential operator associated with
the elliptic integral $E$.
In  Sec. 3 we present an algebraic formulation of the Fuchsian equations 
for $C(N,N)$ by studying the Riccati formulation of solutions to 
PVI  for $N=1,2$ which are related to differential structures on
certain rational curves. 
In Sec. 4 we extend our considerations to the discrete generalization
of Painlev\'e VI, namely a quadratic double recursion on the two-point
correlation functions $\, C(N, \, M)$
together with their dual $\,C^{*}(N, \, M)$. We will show that these
structures can be generalized, mutatis mutandis, to the  $\, C(N, \, M)$'s.
The $\, C(N, \, M)$'s are also solutions of Fuchsian linear ODE's,
with a quadratic increasing order.
The associated differential operators are now homomorphic to direct sums
of $N$-th symmetric  power of the second order differential operator
associated with the complete elliptic integral $\, E$.
The $\, C(N, \, M)$'s are actually sums of several
homogeneous polynomials in the complete elliptic integrals $\, E$ and  $\, K$.
This is a consequence of various remarkable simplifications in the ``discrete
Painlev\'e'' double recursions,
like the fact that algebraic or rational expressions become polynomials by
remarkable factorizations and by the occurrence of perfect squares.
Combining these various results together, one has some quite curious 
and fascinating alchemical wedding between 
complete elliptic integrals, rational curves and discrete generalizations
of Painlev\'e VI (and Hirota-B\"acklund transformations). 
The confrontation between the non-linear
Painlev\'e world and the linear Fuchsian world (Painlev\'e versus Fuchs) 
yields the emergence of quite interesting
structures of differential nature but also of {\em algebraic geometry} nature.
We finally see  in Sec. 5 that, in the case of the $\, C(N, \, N)$ 
holonomic solutions, 
 the $\, p$ and $\, q$  Malmquist's variables
corresponding to the Hamiltonian structure of the sigma form of Painlev\'e 
VI are remarkably {\em rational expressions} of $\, E$ and  $\, K$, and even 
{\em rational expressions} of  $\, E/K$. We have the same result for the
$\, \sigma$ and $\, \sigma'$ variables.  
These last results are in complete agreement
with the previous mentioned results, namely the rational character
of the algebraic curves corresponding to the existence of  
holonomic solutions $\, C(N, \, N)$'s
for the sigma form of Painlev\'e VI, and 
the existence of simple Riccati equations
for the uniformizing parameter. 

The number of new exact results we have obtained being quite large 
and the explicit formulas for some of these results being quite cumbersome, 
we will just sketch here these new exact results, giving the simplest formulas.
More exhaustive formulas will be given in forthcoming publications.

\section{Solutions of sigma form of Painlev\'e VI and Fuchsian linear ODE's}
\label{first}

We consider, from now on,  the isotropic square Ising model and the high
temperature regime, i.e., $t=s^4$ where $s=\sinh(2J/kT)$. The introduction of these two variables,
 $t$ and $\, s$, may look a bit redundant: the  variable $\, t$ is well-suited to
write down  our results on diagonal correlations functions, while
the variable $\, s$ is clearly better suited for non diagonal correlations.
The results for the low temperature regime are similar.
The diagonal two-point correlation functions of the square Ising model
$C(N,N)$ and its dual $C^*(N,N)$ can be
calculated from Toeplitz determinants \cite{kau-ons-49, mo-po-wa-63, mcc-wu-73}:
\begin{eqnarray}
\label{cnn}
C(N,N) =\,  {\rm det} \Bigl(      a_{i-j}       \Bigr), \qquad    1 \le i,\, j \le N       \\
C^*(N,N) =\, (-1)^{N} {\rm det} \Bigl(      a_{i-j-1}       \Bigr), 
\qquad    1 \le i,\, j \le N       
\end{eqnarray}
where the $a_n$'s read  in terms of ${_2}F_1$ hypergeometric function
\begin{eqnarray}
\label{lesan}
&&a_n  = - {\frac{(-1/2)_{n+1}}{(n+1)!}} \,
 t^{n/2+1/2}\cdot {_2}F_1 \Bigl( 1/2, n+1/2; n+2; t \Bigr),\,\,\,
 n \ge -1
 \\
&&a_n  = - {\frac{(1/2)_{-n-1}}{(-n-1)!}} \, 
t^{-n/2-1/2}\cdot {_2}F_1 \Bigl(- 1/2,- n-1/2;- n; t \Bigr),\,\,\,n
 \le -1 \nonumber
\end{eqnarray}
where $(\alpha)_n$ is the usual Pochhammer symbol.

The diagonal two-point correlation functions of the square Ising model
$\, C(N,N)$ and $\, C^{*}(N,N)$ being given 
by the Toeplitz
 determinant (\ref{cnn}) whose entries are solution
of {\em linear} second order differential equations,
they are {\em necessarily} solutions of a {\em linear} differential equation,
with order $\, N! \cdot 2^N$ as an upper bound for generic entries of
the determinant.

Since the diagonal two-point correlation functions of the square Ising model
$\, C(N,N)$ are given by the determinants (\ref{cnn}), it is straightforward
to obtain a sufficiently large number of series coefficients and to
get the linear differential equations satisfied by
 these series~\cite{ze-bo-ha-ma-04,ze-bo-ha-ma-05,ze-bo-ha-ma-05b}.
Denoting by $\,D_t$ the derivative with respect to the 
variable $\,t$, 
the first  linear differential operators $\,L_{NN}$  corresponding to
the $\, C(N,N)$ are
\begin{eqnarray}
\label{Fuchs}
&& L_{11} \, = \, \, D^2_t+{\frac {1}{t}} \cdot  D_t \,
 +\, {{1} \over {4}} \,{\frac {1}{ \left( t-1 \right) {t}^{2}}}, 
 \label{fe1} \\
&& L_{22} \, = \, \,  D^3_t\, 
+2\,{\frac { (t-2) }{ \left( t-1 \right) t}}\cdot D^2_t
-{\frac {1}{
 \left( t-1 \right) {t}^{2}}} \cdot D_t \, 
 - {{1} \over{2}}\,{\frac {t+2}{{t}^{3} \left( t-1 \right)^{2}}},\label{fe2}\\
&& L_{33} \, = \, \,
D^4_t\, +2\,{\frac { \left( t-5 \right)  }
{ \left( t-1 \right) t}}\cdot D^3_t 
\, +{{1} \over {2}} \,{\frac { \left(
41  -11\,t -2\,{t}^{2} \right) }{{t}^{2} 
\left( t-1 \right)^{2}}}\cdot D^2_t   \nonumber \\
&&\qquad  +{{1} \over {2}} \,{\frac { \left( 2\,{t}^{2}+2\,t-5 \right) 
}{{t}^{3} \left( t-1 \right) ^{2}}}\cdot  D_t
+{\frac {9}{16}}\,{\frac {15 +13\,t +4\,{t}^{2}}
{ \left( t-1 \right)^{3} {t}^{4}  }},  \label{fe3}\\
&& L_{44} \, = \, \,Dt^{5}\, -\,{\frac {20}{t \left( t-1 \right) }} \cdot  Dt^{4} \, 
+{\frac { \left(113\, +7\,t\, -2\,{t}^{2} \right) }{ \left( t-1 \right)^{2}{t}^{2}}}\cdot Dt^{3}   \, 
\nonumber \\
&&\qquad  -{{1}\over {2}}\,{\frac { \left(322\, +95\,t-9\,{t}^{2}-16\,{t}^{3} \right) }
{ \left( t-1 \right)^{3}{t}^{3}}} \cdot Dt^{2} \, \\
&&\qquad 
+{\frac { \left(97\, +40\,t-10\,{t}^{2}-12\,{t}^{3} \right) }{ \left( t-1\right)^{3}{t}^{4}}} \cdot Dt\, 
-4\,{\frac {32+33\,t+20\,{t}^{2}+5\,{t}^{3}}{ \left( t-1 \right)^{4}{t}^{5}}}, \nonumber \\
&& L_{55} \, = \, \,Dt^{6}\, -5\,{\frac { \left( t+7 \right) }{t \left( t-1 \right) }}\cdot   Dt^{5} \, 
+ {{1}\over {4}}\,{\frac { \left( 52\,{t}^{2}+483\,t+1617 \right) }
{ \left( t-1 \right) ^{2}{t}^{2}}}\cdot  Dt^{4}  \, \nonumber \\
&&\qquad 
-{{1}\over {2}}\,{\frac { \left( 4\,{t}^{3}+370\,{t}^{2}+1707\,t+3503 \right) }
{ \left(-1+t \right)^{3}{t}^{3}}}\cdot Dt^{3} \,  \\
&&\qquad 
-{{1}\over {16}}\,{\frac { \left( 1552\,{t}^{4}-1016\,{t}^{3}-13191\,{t}^{2}-29618\,t-29855 \right) }
{{t}^{4} \left( t-1 \right)^{4}}}\cdot  Dt^{2}  \, \nonumber \\
&&\qquad 
+{\frac {5}{16}}\,{\frac { \left( 720\,{t}^{4}+640\,{t}^{3}-2175\,{t}^{2}-6912\,t-8801 \right)}
{ \left(  t-1 \right)^{4}{t}^{5}}} \cdot Dt\, \nonumber \\
&&\qquad 
+{\frac {25}{64}}\,{\frac {784\,{t}^{4}+  3428\,{t}^{3}+6921\,{t}^{2}+8650\,t+7865}
{ \left( t-1 \right)^{5}\, {t}^{6} }},
\nonumber \\
&&  L_{66} \, = \, \,D^7_t\, 
-14\,{\frac { \left( 4+t \right)}{ \left( t-1 \right) t}} 
\cdot D^6_t\, 
+14\,{\frac { \left(
81+39\,t+7\,{t}^{2} \right)}{{t}^{2} \left( t-1 \right) ^{2}}}
\cdot D^5_t\nonumber \\
&&\quad \quad -{\frac { N_4 }{ \left( t-1 \right) ^{3}{t}^{3}}}
\cdot D^4_t\, \, 
+{\frac { N_3 }{{t}^{4} \left( t-1 \right) ^{4}}}\cdot D^3_t \, \, 
+{\frac { N_2 }{ \left( t-1 \right) ^{5}{t}^{5}}}\cdot  D^2_t\nonumber \\
&&\quad\quad  -{{1} \over{4}}\,{\frac {  N_1 }{{t}^{6} 
\left( t-1 \right) ^{5}}}
 \cdot D_t\, \, 
-{{9} \over{2}}\,{\frac {  N_0    }{ \left( t-1 \right)^{6}{t}^{7}}}, 
\label{fe6}
\end{eqnarray}
where
\begin{eqnarray}
&& N_4\, = \, 10162+7059\,t \, +2411\,{t}^{2}+376\,{t}^{3}, \quad     \nonumber \\
&& N_3\, = \,  37973 +35162\,t+17893\,{t}^{2}+5116\,{t}^{3}+500\,{t}^{4}, \nonumber \\
&& N_2 \, = \, -28706-55327\,t-46180\,{t}^{2}-21437\,{t}^{3}
 -3358\,{t}^{4}+1736\,{t}^{5}, \nonumber \\
&& N_1 \, = \,   -390548-402496\,t-240997\,{t}^{2}-63239\,{t}^{3} \nonumber \\
&& \qquad +24152\,{t}^{4}+25088\,{t}^{5},  \\
&& N_0 \, = \, 23814+   26839\,t+24583\,{t}^{2}+16599\,{t}^{3} 
+7345\,{t}^{4}+1620\,{t}^{5}  \nonumber
\end{eqnarray}
These operators are of order $N+1$ and are irreducible.
We further note that, in contrast to the Fuchsian equations 
for the $n$-particle contributions $ \chi^{(n)}$'s of the susceptibility of
the Ising model
\cite{ze-bo-ha-ma-04,ze-bo-ha-ma-05,ze-bo-ha-ma-05b,ze-bo-ha-ma-05c},
the Fuchsian differential equations 
satisfied by the $\,C(N,N)$'s have {\em no apparent singularities}.
The linear differential operators, $\, L^{*}_{NN}$,   
for the $\,C^{*}(N,N)$'s are obtained by the change 
 $\, t  $  into  $\, 1/t  $ 
in the previous differential operators.

These {\em Fuchsian} differential equations 
(\ref{fe1}-\ref{fe6}) for the $\,C(N,N)$'s have
the following general form :
\begin{equation}
 \sum_{i=2}^{N+1} t^{i} (t-1)^{i-1} P_{i}^{(N)}(t)
 \cdot  D_t^{i}\,\, 
  +\,\,t(t-1)\, P_1^{(N)}(t) \cdot  D_t
 + P_0^{(N)}(t)\, =\, \, 0  \quad \quad 
\label{gform}
\end{equation}
where $P_i^{(N)}(t)$ is a polynomial in $t$ of degree 
$N+1-i$  for $i=2,\cdots,N+1$ and $ P_1^{(N)}(t)$ and $P_0^{(N)}(t)$ 
are of degree $N-1$.
  
The only singular points of (\ref{gform}) are the three regular singular points
$\, t \, =\, 0, \, \infty, \, 1  $.
From the indicial equation of the differential equations for the first
$L_{NN}$'s, 
we infer the remarkably simple expressions of the critical exponents
$\rho^{(1)}$, $\rho^{(\infty)}$ and $\rho^{(0)}$ at respectively the regular
singular points $t=1$, $t=\infty$ and $t=0$
\begin{eqnarray}
\rho_n^{(1)} &=&\, (n-1)^2  \\
\rho_n^{(\infty)} &=&\, {\frac{5}{8}}+{\frac{3}{4}}N 
+{\frac{1}{4}}n^2-{\frac{1}{4}}(2N+3)\cdot n
-{\frac{(-1)^n}{4}}n +{\frac{(-1)^n}{8}}(2N+3)   \nonumber \\
\rho_n^{(0)} \, &=&\,  -{\frac{1}{8}}+ {\frac{3}{4}} N\, 
 +{\frac{1}{4}}\, (n+1)(n+2) \,  \\
&& \quad \quad -{\frac{1}{2}}\, (N+3) \cdot n \,  
+{\frac{(-1)^n}{4}}(n+1)\,  -{\frac{(-1)^n}{8}}(2N+5) \nonumber 
\end{eqnarray}
where $n=1,2, \cdots N+1$.

\subsection{Local solutions at $t=0,1,\infty$}
\label{Formal}

It is of interest to compare the local expansion (\ref{expand}) of 
the PVI equation with the exponents of the Fuchsian
equations\footnote{Recall that, for Ising case and for $T >T_c$, 
$\tau =t^{1/4}\,C_N$ (resp.  $\tau =\,C_N$ for $T <T_c$).}.
For concreteness we concentrate on $t=1^{-}$ which corresponds to $T=T_c^{+}$ in
the Ising model. We have the following coefficients
in (\ref{expand}) valid for $0<\alpha<1$
\begin{eqnarray}
&& a_1(0,0;\alpha)= 1, \\
&&a_1(0,-1;\alpha)=a_1(0,1;-\alpha)
\nonumber\\
&& ={1\over 16 \alpha^2(1-\alpha)^2} (\alpha-v_1-v_2-v_3+v_4)(\alpha-v_1-v_2+v_3-v_4) \times
\nonumber \label{a01}\\
&& \qquad  (\alpha-v_1+v_2-v_3-v_4)(\alpha+v_1-v_2-v_3-v_4),\\
&&a_1(1,0;\alpha)=-{\alpha^2\over 8}+{1\over 2}
(-v_1v_2+v_1v_3+v_1v_4+v_2v_3+v_2v_4-v_3v_4) \nonumber\\
&&+{1\over 8\alpha^2} \; (v_1+v_2+v_3-v_4)(v_1+v_2-v_3+v_4) \times \nonumber\\
&&\qquad \qquad \qquad  (v_1-v_2+v_3+v_4)(v_1-v_2-v_3-v_4),    \\
&&a_1(0,-2;\alpha)=\, a_1(0,2;-\alpha)={a_1(0,-1;\alpha)^2\over
  265(\alpha-1)^2(\alpha-2)^4(\alpha-3)^2}\label{a02} \\
&&\times[(\alpha-2)^2-(v_1+v_2+v_3-v_4)^2]
[(\alpha-2)^2-(v_1+v_2-v_3+v_4)^2]\nonumber\\
&&\times[(\alpha-2)^2-(v_1-v_2+v_3+v_4)^2][(\alpha-2)^2-(-v_1+v_2+v_3+v_4)^2] 
\nonumber 
\end{eqnarray}

For the Ising case (\ref{isingv}) this reduces to
\begin{eqnarray}
\label{thepIsing}
&&p_1=\alpha^2/4,   \\
&&a_1(0,-1;\alpha)=a_1(0,1;-\alpha)=\, {\alpha-2N\over 16\alpha}, \quad
a_1(1,0;\alpha)={(1- \alpha^2)\over 8}, \\
&&a_1(0,-2;\alpha)=a_1(0,2;-\alpha)=
{a_1(0,-1;\alpha)^2 \cdot \bigl( (\alpha-2)^2-(2N)^2 \bigr) \over 256(\alpha-2)^2}
\label{a102}
\end{eqnarray}

When used in (\ref{expand}) these expressions will reproduce the $N+1$
exponents of $L_{NN}$ at $t=1$  where, in the limit
$\alpha\rightarrow 0$, the terms in (\ref{expand}) with $x^{k^2\pm
  k\alpha}$ become $(t-1)^{k^2}\ln^k(t-1).$ 
We see from
(\ref{a102}) that, when $\alpha=0$, $a_1(n,\pm2;0)=0$ for $N=1$
which is consistent with the fact that $C(1,1)$ satisfies a second
order linear differential equation. We have carried the expansion  to order 
$(t-1)^{12}$. In particular we have obtained the coefficient of 
$(t-1)^9\ln^3|t-1|$ and have verified that it vanishes for $N=1,2$
and have obtained all terms in the expansion of $C(N,N)$ given
in \cite{or-ni-gu-pe-01b}.

More generally the conditions that 
there exists a value of
$\alpha$ such that $a_j(n,k;\alpha)=0$ for all $k$ sufficiently large
is a condition necessary for $\tau$ function of the PVI equation to
satisfy a linear differential equation of finite order and the series 
\begin{equation}
\sum_{n=0}^{\infty}\, a_j(n,k;\alpha) \cdot x^{k^2+k\alpha+p_j+n}
\end{equation}
will be solutions to the Fuchsian equation. 
For example one condition for a second order Fuchsian equation 
is $a_1(0,1;\alpha)=0,\,\, a_1(0,-2;\alpha)=0$, 
which are satisfied if, respectively,
\begin{eqnarray}
\label{cond3}
\alpha=\, -v_1-v_2+v_3-v_4, \qquad  \alpha-2=\, -v_1+v_2-v_3-v_4
\end{eqnarray}
implying $\, v_2-v_3\,=\,-1$, 
which is the restriction Forrester and  Witte \cite{Forrest} needed for a
solution of PVI to satisfy a hypergeometric equation. This condition 
implies that the $\, \tau$ functions  are determinants of hypergeometric functions.
We thus see that, at order $x^{p+4+(1-\alpha)}$, the local expansion provides a
  necessary condition for the reduction of a one parameter family of
  solutions to PVI to a solution of a second order linear differential  equation. 
By examining the vanishing of $a(0,k;\alpha)$ for higher values of
$k$ necessary conditions for the existence of one parameter families
satisfying higher order linear differential  equations will be obtained.
Similar necessary conditions can be obtained from the local expansions
at $t=0,\infty.$

\subsection{The Fuchsian differential operators as $\, N$-th symmetric power}
\label{power}

The most profound and surprising structure of the solutions of PVI
which satisfy Fuchsian equations is, however, not seen in 
these local expansions and, thus, it is important to observe that the
operators $\, L_{NN}$ given in (\ref{fe1}-\ref{fe6}) for $C(N,N)$ have the
remarkable property that they are equivalent 
\footnote[8]{For the equivalence of differential
 operators see (e.g.) \cite{Singer,hoeij2,PutSinger}.} to 
the $\, N$-th symmetric power\footnote[9]{For the definition
of the symmetric power of a differential
 operators see (e.g.) \cite{PutSinger,SingerUlmer,HMW}.} of  $\, L_{11}$: 
\begin{eqnarray} \label{relation32}
\label{Sym}
  A_N \cdot  L_{NN} \, = \, \, Sym^N(L_{11}) \cdot R_N 
\end{eqnarray}
The first  $\, A_N$ and  $\, R_N$ intertwinners read for $\,N=2$ :
\begin{eqnarray}
 && A_2 \, = \, \,{t}^{2} D_t^{2}\, 
+{{1} \over {4}}\,{\frac {\left( 31\,t-23 \right)\, t }{t-1}} 
\cdot  D_t   \, \, 
+{{3} \over {4}}\,{\frac {15\,t-7}{t-1}}
\\
 && R_2 \, = \, \, {t}^{2} \cdot  D_t^{2}\, +{{3} \over {4}}\,t 
\cdot   D_t\, \,
 -{{1} \over {4}}\,{\frac {3\,t-5}{t-1}} 
\end{eqnarray}
We have calculated exactly these intertwinners up
to $\, N=\, 6$ but the
expressions are too large to be given here.
As a consequence of this  property (\ref{Sym}) the differential
Galois group of $L_{NN}$ is not a $\, SL(N+1, C) $ group as we could expect
at first sight, but an $\, SL(2, C) $ group 
in the symmetric power representation. We expect that this property
extends much more generally to other solutions 
of the general four parameters PVI which satisfy Fuchsian equations. 

Another consequence of (\ref{Sym}) is that the solutions of this order $\, N+1$ differential operator
$\,  L_{NN} $ are actually homogeneous polynomials of degree $\, N$ in the two solutions 
of $\, L_{11}$, see~\cite{PutSinger,SingerUlmer,HMW}. 

Let us now introduce the two 
 elliptic integrals
\begin{eqnarray}
K\,=\,\, {_2}F_1 \left( 1/2, 1/2; 1; s^4 \right), \quad \quad
E\,=\,\, {_2}F_1 \left( 1/2, -1/2; 1; s^4 \right)
\end{eqnarray}
and the second order linear differential operator for $\, E$ ($D_s$ denotes 
the derivative with respect to $\, s$):
\begin{eqnarray}
\label{Lee}
&& L_{E} \, = \, \, \, D_s^{2}\,
 +{\frac {D_s}{s}}\, -4\,{\frac {{s}^{2}}{{s}^{4}-1}}            
\end{eqnarray}
This operator actually identifies with $\, L_{11}^{*}$.

One can easily show that the  second order linear differential operator
$\, L_{11}$ (associated with $\, C(1, \, 1)$ and written in the variable $s$)
and the second order linear differential operator $L_E$ are equivalent :
\begin{eqnarray}
\Bigl( {\frac{s^4-1}{s}}\cdot  D_s \, +6s^2 \Bigr) \cdot L_{11} \, 
= \,\, \,  L_E \cdot
\Bigl( {\frac{s^4-1}{s}}\cdot  D_s\, -2/s^2 \Bigr)
\end{eqnarray}
More generally one can show
in the $\, s$ variable, that the $\, L_{NN}$'s  are actually 
equivalent to the   $\, L_{NN}^{*}$' s.
Since $\, K$ can be simply expressed in terms of $E$ and its first derivative,
the $C_{N,N}$'s are thus solutions of an operator which is homomorphic
to $Sym^N(L_E)$ :
\begin{eqnarray}
\label{Sym21}
&&  \tilde{A}_N \cdot  L_{NN} \, = \, \,
 Sym^N(L_{E}) \cdot \tilde{R}_N, \qquad \quad \hbox{or:}  \\
\label{Sym2}
&&  L_{NN} \cdot B_N \, = \, \,  S_N  \cdot Sym^N(L_{E}) 
\end{eqnarray}
where the intertwinners $\, B_N$ and  $\, S_N$ (or $\, \tilde{A}_N$
and  $\,\tilde{R}_N$)   are linear differential
operators of order $\, N$. In fact, beyond  $C(N,N)$, relations 
(\ref{relation32}), (\ref{Sym21}), (\ref{Sym2})
 relate {\it all} solutions of $L_{NN}$ to
$Sym^N(L_E)$.
From (\ref{Sym2}) one can easily deduce that 
the diagonal two-point correlation functions
$\, C(N, \, N)$ can be deduced as the action of a linear differential operator of order $\, N$
  on the $\, N$-th power of the complete elliptic E :
\begin{eqnarray}
\label{cnnsn}
C(N, \, N) \,  \, = \, \,  \,  B_N (E^N) 
\end{eqnarray}

The expressions of the intertwinners $B_N$ can also be retrieved from the determinental
 expressions of the C(N,N) and the relations between the hypergeometric functions $a_n$,
 $a_{n-1}$ (see (\ref{cnn}), (\ref{lesan})) and its derivative. This gives a general method
 to obtain the differential operators $\, L_{N,N}$.

\subsection{The $\, C(N, \, N)$'s as homogeneous polynomial of the complete
elliptic integrals $\, E$ and $\, K$}
\label{homog}

The property (\ref{Sym}), or (\ref{Sym2})
 can be illustrated by considering the specific
solution $\, C(N, \, N)$ of the $N+1$ order differential equations $L_{NN}$.
The matrix elements $a_n$ of the Toeplitz determinant representation may
all be expressed as linear combinations of the
 elliptic integrals $\, E$ and $\, K$,
and, thus, $C(N,N)$ will be given as polynomials in these functions
and this is in agreement with the previous relation (\ref{cnnsn}).
For low orders these polynomials have been presented by Ghosh and
Shrock \cite{Ghosh1}. For example
\begin{eqnarray}
\label{c33expl}
&& C(2, \, 2) \, = \, {{1} \over {3\, s^4}} \cdot
 \Bigl( 3\, \left( {s}^{4}-1 \right)^{2 \cdot }K^{2}
\, +8\, \left( {s}^{4}-1 \right) \cdot E K \, 
- \left({s}^{4} -5 \right) \cdot  E^{2} \Bigr)\nonumber \\
 && C(3, \, 3) \, = \,{{4} \over {135 \, s^{10}}} 
\cdot P_3(E, \, K), \quad \quad \quad \hbox{where:} \quad \quad 
P_3(E, \, K)\, = \,\nonumber \\
 && \quad \quad \left( 33\,{s}^{4}-1 \right) 
 \left( {s}^{4}-1 \right)^{3} \cdot K^{3}\,
 +3\, \left( {s}^{8}+48\,{s}^{4}-1 \right)  
\left( {s}^{4}-1 \right)^{2} \cdot E \,K^{2} \,\nonumber \\
 && \quad \quad   -3\, \left({s}^{4}-1 \right) 
 \left( {s}^{12}+3\,{s}^{8}-69\,{s}^{4}+1 \right)\cdot E^{2}\, K\, \\
 && \quad \quad  
 - \left(1+21\,{s}^{8}-96\,{s}^{4}+10\,{s}^{12} \right) \cdot E^{3}
\nonumber
\end{eqnarray}
We note that these expressions are respectively quadratic and cubic
{\em homogeneous} polynomial in  $\, E$ and $\, K$.
We have obtained similar expressions for  all the $\, C(N, \, N)$
and $\, C^{*}(N, \, N)$ for $\, N\, = \, 4, 5, \, 6, \, \cdots, \, 21$, 
and relation (\ref{cnnsn}) gives similar
relations for {\em any} values of $N$.
They are {\em homogeneous} polynomial of degree $\, N$ in the complete
elliptic integrals\footnote[5]{This result can also be found 
in the Eqs. (2.16)-(2.19) of~\cite{Odyssey1}, which also show 
very explicitly  that
$C(N,N)$  is a homogeneous polynomial of $\, E$ and $\, K$ of degree $\, N$ for all $\, N$
(something that is already, albeit less explicit, in the appendix
of Montroll, Potts, and Ward~\cite{mo-po-wa-63}.).
} $\, E$ and $\, K$, with simple rational coefficients
(a polynomial in $\,s$ with integer coefficients  divided by 
some power of $\,s$). 
From a physics viewpoint one should note that the particular 
 rational coefficients one gets in front of 
the monomials $\, E^k \cdot K^{N-k}$,
are far from being arbitrary as a general formula like (\ref{cnnsn})
could suggest.  These coefficients are
such that, for instance, the linear differential 
equation for the $\, C(N, \, N)$'s
has no apparent singularities.  Furthermore, the contribution associated
to the various monomials  $\, E^k \cdot K^{N-k}$ clearly have poles ($s^{-10}$ 
or $s^{-4}$ in the previous example (\ref{c33expl})).  These coefficients are
also  ``fined-tuned'' in such a way that, for
instance, these various poles cancel together, in order to give an expression
with a well-defined high-temperature series expansion (series at $\, s=\, 0$). 
 We have many other remarkable properties corresponding to the 
behavior of the $\, C(N, \, N)$'s near $\, s=1$ or $\, s \, = \, \infty$.

\subsection{Non-trivial disentangling of solutions of linear Fuchsian ODE's near $\, t=0$.}
\label{nontriv}
Let us make here a  comment on the existence of 
 surprisingly simple hypergeometric solutions
of the $\, N$-dependent sigma form (\ref{jimbo-miwa}) of Painlev\'e VI.
Consider the second order differential operator:
\begin{eqnarray}
\label{Ljm}
L_{h}\, = \, \, D_t^{2}\, 
+ \left( {{1} \over {t}} \, +\, {{1} 
\over {2 \, \left( t-1 \right)}}  \right)\cdot   D_t\,\,  
-{{1} \over {4}} \,{\frac {{N}^{2}}{{t}^{2}}}
+{\frac {1}{16 \left( t-1 \right)^2}}\,
\end{eqnarray}
which has regular singularities at
$t=0$, $t=1$ and $t=\infty$ with respectively
the critical exponents ($\pm N/2$), ($1/4, 1/4$) and ($1/4 \pm N/2$).

It can be verified that {\em any linear combination} of the two 
solutions of (\ref{Ljm})  satisfies the $\, N$-dependent sigma form (\ref{jimbo-miwa})
Painlev\'e VI equation  {\em for arbitrary} $\, N$,
 not {\em necessarily an integer}. For instance,
when $\, N$ is {\em not an integer}, one has the two following solutions of (\ref{jimbo-miwa}):
\begin{equation}
\sigma\,=\,\, t\, (t-1)\, {d\ln \tau \over dt}\,-{{1} \over {4}} \qquad 
\hbox{where :} \qquad \tau\,=\,\, f_{+}\, +\lambda \cdot  f_{-}
\end{equation}
where $\, f_{\pm}$ are the two independent solutions of (\ref{Ljm}) :
\begin{equation}
f_{\pm}\, =\, 
t^{\pm N/2} \cdot  (1-t)^{1/4} \cdot  {_2}F_1( [1/2, 1/2 \, \pm\, N],\, [1 \, \pm\, N], t)
\label{sol1}
\end{equation}
When the parameter $\, N$ is an integer (and only in this case), that is to 
say in the Ising case we are interested
in, the second order differential
 operator $\, L_h$ is, after conjugation by $\, (1+s^2)^{1/2}$, 
{\em equivalent to} $\, L_E$; when $\, N$ is an integer,
 one solution is given above 
in term of a hypergeometric function analytic at $\, t=0$, and the other
 one has a logarithmic singularity 
at $\, t=0$ (and similarly for $\, t=1$ and $\, t=\infty$).

At first sight the existence of such ``additional'' solutions 
should not be seen as a surprise:
we certainly expect the solutions of the $\, N$-dependent sigma
form of Painlev\'e VI that are also, at the same time, solutions
of a {linear} (Fuchsian)
ODE, to be a quite complicated ``stratified'' space.
However, let us focus on the series expansion at $\, t=0$ of the analytic
solution of (\ref{Ljm}), which simply reads
\begin{eqnarray}
\label{hyper}
&&h_N  \, = \, \, {\frac{1}{4^N}} {\frac{\Gamma(2N+1)}
{\Gamma(N+1)^2}}\cdot f_{+} \nonumber  \\
&&\quad \, = \, \,\, c_0(N) \cdot t^{N/2} + c_1(N)\cdot t^{N/2+1}+
c_2(N)\cdot t^{N/2+2} + \cdots 
\end{eqnarray}
The coefficients $\,c_k(N)$  in the series expansion of (\ref{hyper}) read :
\begin{eqnarray}
\label{ck}
&&c_k(N)\,\,\,   =\,    \\
&&{\frac{1}{4^N}} {\frac{\Gamma(2N+1)}{\Gamma(N+1)^2}}\,
{\frac{\left(-1/4 \right)_k}{k!}}\,
 {_3}F_2\left([1/2, 1/2+N, -k],[1+N, 5/4-k],1    \right) \nonumber
\end{eqnarray}

Let us now consider the series expansion of 
the diagonal correlation functions $C(N,N)$:
\begin{eqnarray}
C(N,N) =\,  d_0(N) \cdot t^{N/2} + d_1(N)\cdot t^{N/2+1}+
d_2(N)\cdot t^{N/2+2} + \cdots
\end{eqnarray}
where $\,d_0(N)$,$\,d_1(N)$ and $\,d_2(N)$ read respectively  :
\begin{eqnarray}
&&{\frac{\Gamma(2N+1)}{\Gamma(N+1) \Gamma(N+1)}}\, {\frac{1}{4^{N}}}, 
\quad \quad
 {\frac{\Gamma(2N+1)}{\Gamma(N+1)\Gamma(N+2)}}\, {\frac{N}{4^{N+1}}}, \,\,  \cdots  
\end{eqnarray}
One has the following  result, that may look quite surprising
at first sight:
the coefficients $\,c_k(N)$ of the solution (\ref{hyper})
and the coefficients $\,d_k(N)$ of the diagonal 
two-point correlation functions $C(N,N)$,
solution of the order $\, N+1$  Fuchsian  ODE 
are {\em identical up to} $\, k \, = \,3N/2+1$:
\begin{eqnarray}
\label{boundary}
C(N,N)\, - \,\, h_N \, \, =\,\, \, {{1} \over {16}} \Bigl( {{(1/2)_N \cdot ((3/2)_N)^2 } \over {
 \Gamma(N+2) \Gamma(N+3)^2 }}    
 \Bigr) \cdot t^{3N/2+2} \, + \, \, \cdots 
\end{eqnarray}
The coefficient in (\ref{boundary}) in front of  $\, t^{3N/2+2}$
can be seen as the initial condition defining\footnote[3]{For $\, T<T_c$, one has 
$\, C(N,N)\, -\,\, (1-t)^{1/4} \,\, =\,\, \, 
1/4 \cdot((1/2)_N \, (3/2)_N)/((N+1)!)^2 \cdot t^{N+1} \, + \cdots $, the  coefficient
in front of  $\, t^{N+1}$ corresponding to the initial condition defining $\, C(N,N)$
in the low-temperature regime~\cite{Ghoshalone}.
} $\, C(N,N)$. 
Seeking  for conditions allowing solutions of the sigma form of Painlev\'e VI
 to be also (the log-derivative of) solutions of linear Fuchsian differential equations, 
this difficulty to disentangle, near $\, t=0$, a solution of a second order differential equation
and a solution of linear Fuchsian differential equations of arbitrary $\, N+1$ order,
seems to indicate that series analysis like (\ref{expand})
may not be the easiest approach to take into account such subtle\footnote[8]{Cauchy's theorem 
does not apply  to PVI at t=0 or t=1. As a consequence, even with given boundary  
conditions (a large set of first terms in the series), there can be ``branching''   
in the series computation. These subtle ``branching''  
series calculations will be adressed elsewhere.} fine-tuning: we 
need a less analytical and   more ``global'' {\em algebraic} approach.

\section{Algebraic viewpoint of the Fuchsian differential equations}
\label{conditions}
The existence of $C(N,N)$ as solutions common to the sigma
 form of Painlev\'e VI
equation and to linear Fuchsian differential equations 
can be addressed on an effective
algebraic geometry approach of differential equations as introduced
explicitly by J.F. Ritt \cite{Ritt,Ritt1}. 
This approach amounts, when working with various linear and non-linear
differential equations, to introducing as many variables as the number of
derivatives of the function we study. 
The analysis of the compatibility between these various linear and
non-linear differential equations will correspond 
to considering an algebraic variety given by various polynomial
relations on these variables.
These relations can be studied from the algebraic viewpoint 
(parametrization when the genus is zero or one, 
birational transformations\footnote[5]{At this step
it is worth recalling that B\"acklund transformations 
are actually {\em birational transformations} in ``some''variables.},
singularity analysis, blow-up, etc.).
The very last step, recalling that the various introduced variables
are not independent but can be deduced from each other by successive
derivation, provides further constraints.
In other words a set of differential equations is seen as an algebraic
variety plus some differential structure on top of it. 

Let us show how this algebraic viewpoint of differential equations works
in our (subtle) compatibility problem of the sigma form of Painlev\'e VI
and the Fuchsian linear ODE's of arbitrary order $\, N+1$.
The correlation function $\, C(1, \, 1)$ satisfies a second order linear
differential equation which can be written in a Riccati form in terms of
$ \sigma(t)$ and $ \sigma'(t)$.
More generally, the $\, N+1$ order Fuchsian linear ODE
satisfied by the $\, C(N, \, N)$'s can be written in a
``generalized Riccati form~\cite{Ince,Poole}''  in terms of  $ \sigma(t)$,  $ \sigma'(t)$
and its successive derivatives $ \sigma^{(n)}(t)$
up to $\, n\, = \, N$
(where  $ \sigma(t)$ is deduced from $\, C(N, \, N)$
by the logarithmic derivative relation (\ref{sigma-bas})).
Similarly, the sigma form of Painlev\'e VI equation (\ref{jimbo-miwa})
is not seen as a non-linear ODE, but as a polynomial relation between
the three variables $ \sigma(t)$,  $ \sigma'(t)$ and $\,\sigma''(t)$.

Introducing the variables $ S_0\, = \sigma(t)$, $S_1\, = \sigma'(t)$,
$\, S_2\, = \sigma''(t)$, etc., 
the third order Fuchsian linear ODE for  $\, C(2, \, 2)$,
yields a ``generalized Riccati form'' which is a polynomial relation between 
$ S_0$,  $ S_1$ and $S_2$
\begin{eqnarray}
\label{C22S}
&& 64\,{t}^{2}\left (t-1\right )^{2} S_2
-16\,t\left (8\,t+5\right )\left (t-1\right )\cdot  S_1 \nonumber \\
&& 
+192\,t\left (t-1\right )S_0 S_1
+64\,{S_0}^{3}- 16\,\left (16\,t+1\right ){S_0}^{2} \nonumber \\
&& +4\,\left (32\,{t}^{2}+16\,t-21\right ) \, S_0  +45=0
\end{eqnarray}
The elimination of the variable  $S_2$ between this
``generalized Riccati form'' and  (\ref{jimbo-miwa}) seen  as a
polynomial relation between the three variables
$S_0$, $S_1$ and $S_2$ yields an algebraic relation 
between $ S_0=\sigma(t)$ and   $ S_1=\sigma'(t)$ which reads:
\begin{eqnarray}
\label{nappe22}
&& \left (4\,S_0-3\right )\left (64\,{S_0}^{3}
-16\,\left (16\,t+1\right ){S_0}^{2} 
 +4\,\left (64\,{t}^{2}-16\,t-21\right ) \cdot S_0+45\right) \nonumber \\
&& -32\,t\left (4\,S_0-3\right )\left (t-1\right )
\left (8\,t-1-4\,S_0\right )\cdot  S_1 \,\nonumber  \\
&&\quad  +256\,{t}^{2}\left (t-1\right )^{2}{ S_1}^{2}\, =\, \, 0 
\end{eqnarray}
which is compatible with (\ref{C22S}) and (\ref{jimbo-miwa}). This can be
checked by eliminating $S_2$ between the derivative of (\ref{nappe22})
and (\ref{C22S}) or (\ref{jimbo-miwa}) to get again  (\ref{nappe22}).
Or directly by plugging a series expansion or an exact expression 
of $C(2,2)$ in (\ref{nappe22}).

Seen as a relation between $S_0$ and $S_1$
(the variable $\, t$ is considered as a simple parameter),
the algebraic curve (\ref{nappe22}) is actually a {\em rational curve}.
It can thus be parametrized in term of two rational functions:
\begin{eqnarray}
\label{param}
&&S_0 \, = \, \,  \, 
{3 \over 4}\,{\frac { A_{2}\cdot {u}^{2} \, 
+A_{1}\cdot u\, + \, A_{0}}{ B_{2}\cdot {u}^{2}
+ B_{1}\cdot u\, + B_{0}}} 
, \qquad  \\
&&S_1 \, = \, \, \,  {{3} \over {t}} \cdot
 {\frac { \left(  \alpha_1 \cdot u\, +\alpha_0
 \right) \cdot \left( C_{3}\cdot {u}^{3} \, 
+C_{2} \cdot {u}^{2} \, +C_{1}\cdot u \, +
C_{0} \right) }{\left( B_{2}\cdot {u}^{2} \,
 +B_{1}\cdot u + \, B_{0} \right)^{2}}}
\nonumber
\end{eqnarray}
where:
\begin{eqnarray}
&& \alpha_1 \, = \, \,   -6\,t-3+8\,{t}^{2}, \qquad \qquad  \alpha_0 \, 
= \, \, 4\cdot  (1\, -2\, t) \nonumber \\
&&A_0 \, = \, \, -176+48\,t-320\,{t}^{2}+256\,{t}^{3}, \nonumber \\
&&A_1 \, = \, \,120+184\,t-144\,{t}^{2}+768\,{t}^{3}-512\,{t}^{4},  \\
&&A_2 \, = \, \,9-57\,t+24\,{t}^{2}+76\,{t}^{3}
-448\,{t}^{4} +256\,{t}^{5},\nonumber \\
&&B_0 \, = \, \,192\,{t}^{2}-272\,t-112, \quad
B_1 \, = \, \,-8\, \left( 3\,t+1 \right) 
 \left( 16\,{t}^{2}-26\,t-3 \right),\nonumber \\
&&B_2 \, = \, \,45+51\,t -168\,{t}^{2}-260\,{t}^{3}+192\,{t}^{4}, \nonumber \\
&&C_0 \, = \, \,1088+384\,t +2624\,{t}^{2}+1280\,{t}^{3} -1536\,{t}^{4},
 \nonumber \\
&&C_1 \, = \, \,-1296-2816\,t+688\,{t}^{2}
-7776\,{t}^{3} -3840\,{t}^{4}+4608\,{t}^{5} ,\nonumber \\
&&C_2 \, = \, \,108+1848\,t+636\,{t}^{2}
-3328\,{t}^{3}+8304\,{t}^{4} +4416\,{t}^{5} -4608\,{t}^{6}, \nonumber \\
&&C_3 \, = \, \, +189 +36\,t -1323\,{t}^{2} +210\,{t}^{3}
+2460\,{t}^{4}-2792\,{t}^{5} \nonumber \\
&& \qquad  -1856 \,{t}^{6}+1536\,{t}^{7} \nonumber
\end{eqnarray}

In the spirit of the ``algebraic viewpoint of
 differential equations''~\cite{Ritt,Ritt1},
having performed the algebraic geometry calculations we had in mind,
we now recall that there is some differential structure
on this  {\em rational curve} by imposing that the variable $\, S_1 $ is 
actually the derivative with respect to $\, t$ of the  variable $\, S_0$: 
\begin{eqnarray}
\label{recall}
S_1 \, = \, \,  {{d S_0} \over {dt} }  \, = \, \,    { {\partial S_0} \over {\partial u } } 
\cdot   {{d u} \over {dt} }\, 
+ \,  { {\partial S_0} \over {\partial t } } 
\end{eqnarray}
yielding, after some quite nice simplifications, that ${{d u} \over {dt} }$ 
is not a rational expression of $\, u$, as one could expect at first sight, 
but a quadratic polynomial in $\, u$, which gives a simple Riccati form:
\begin{eqnarray}
\label{Ricatti}
&&16\,t\left (t-1\right )\left (6\,{t}^{2}-5\,t-9\right )
\cdot  {\frac{d\,u}{dt}} \, 
 \, \,  = \nonumber \\
&& \left (  63 -135\,t -120\,{t}^{2} -140\,{t}^{3}
+192\,{t}^{4}\right )\cdot {u}^{2}  \\
&& +8\,\left (15 +51\,t +46\,{t}^{2} -60\,{t}^{3}\right )
\cdot  u \, \,  -272-112\,t\, +192\,{t}^{2}
\nonumber
\end{eqnarray}
that can easily be associated with a linear second order differential equation
bearing on some function $\, F$:
\begin{eqnarray}
v \, \, = \, \,  \,  {{ 1} \over {F}} \cdot 
{{d F} \over {dt}} \, \, = \, \, 
-{1 \over 16}\,{\frac {192\,{t}^{4}-140\,{t}^{3}-120\,{t}^{2}-135\,t+63}
{t \left( -1+t \right)  \left( 6\,{t}^{2}-5\,t-9 \right) }}
\cdot u  \qquad \qquad 
\end{eqnarray}

Similar calculations can be performed for $\, N\, = 3$, the generalized Riccati
form for the Fuchsian linear ODE of order four is now a polynomial relation of
the form:
\begin{eqnarray}
\label{Riccati3}
S_3 \, = \, \, \, P(S_0, \, S_1, \, S_2; \, t) 
\end{eqnarray}
where $\, P$ is a polynomial of the three variables $S_0$, 
$S_1$ and $S_2$, the coefficients being rational function
(with integer coefficients) in the variable $\, t$ seen as a parameter.
In order to combine\footnote[3]{Or, in mathematical wording, to calculate
the ideal of these two differential equations.} this generalized Riccati 
form (\ref{Riccati3}) with (\ref{jimbo-miwa}) for $\, N\, =3$, we need,
in order to perform eliminations of variables (ideal of polynomials),
to rewrite (\ref{jimbo-miwa}), the sigma form of Painlev\'e VI taken
for  $\, N\, =3$ as a
relation between $\,\sigma,$ $ \, \sigma'$ and $ \, \sigma''$ and
$ \sigma^{(3)}$ as well. This is easily
obtained by performing the derivative of (\ref{jimbo-miwa}) with respect
to $\, t$, thus getting a polynomial relation
between $\,\sigma,$ $ \, \sigma'$ and $ \, \sigma''$ and $ \sigma^{(3)}$.
Considering this last polynomial relation
and the generalized Riccati form (\ref{Riccati3}), we can easily eliminate
$S_3=\sigma^{(3)}$,
getting a new polynomial relation on $\,S_0$, $ \, S_1$ and
$ \, S_2$. We can, now, eliminate $ \, S_2$ between this new
polynomial relation and (\ref{jimbo-miwa})  for $\, N\, =3$
which is also a polynomial relation on $\,S_0$, $ \, S_1$ and
$ \, S_2$, in order to get, finally, a  polynomial relation on
$S_0=\sigma\, $ and $ S_1= \sigma'$ {\em only}.
This final relation reads:
\begin{eqnarray}
\label{ratioN3}
&&4096\,{t}^{3} \left(t-1 \right)^{3} \cdot S_1^{3}\,
 +256\,{t}^{2} \left(t-1 \right)^{2} \, Q_2  \cdot   S_1^{2}  \\
&& -16\,t \left(t-1 \right)  \, Q_1 \cdot   S_1 \, 
- \left(45 -8\, \left( 2\,t+7 \right)\, S_0 +16\,{S_0}^{2}\right)
 \cdot Q_0 \, \, = \, \, \, 0
\nonumber
\end{eqnarray}
where:
\begin{eqnarray}
&&Q_2 \, = \,  \, 48\,{S_0}^{2}\, -8\, \left( 22\,t+13 \right)\cdot  S_0   \,
 +55 \, +448\,t\,  +64\,{t}^{2} \nonumber \\
&&Q_1 \, = \,  -768\,{S_0}^{4}
+256\, \left( 22\,t+13 \right) {S_0}^{3}-32\, \left( 376\,{
t}^{2}+584\,t+125 \right) {S_0}^{2}\,\nonumber \\
&& \quad +16\, \left( 384\,{t}^{3}+1984\,{t}^{2}+766\,t+25 \right)\, S_0 \, 
+1125+2880\,t-25920\,{t}^{2} \nonumber \\
&&Q_0 \, = \,1575 +16\, \left( 576\,{t}^{3}
-110\,t-145-96\,{t}^{2} \right) S_0  \\
&& \quad -32\, \left( 56\,t-9+264\,{t}^{2} \right) {S_0}^{2}
 +256\, \left( 10\,t+3 \right) {S_0}^{3}  -256\,{S_0}^{4}\nonumber
\end{eqnarray}

Similar calculations (of ideal of differential equations
seen as ideal of polynomials), can be performed, mutatis mutandis,
for $\, N\, = 4, \, 5 \, $   and $ \, 6$.
These eliminations yield polynomial relations in $\, t$, $S_0=\sigma$
and $S_1=\sigma'$ of the form:
\begin{eqnarray}
\label{ratioN}
 \sum_{i\, = 0}^{i=N} \, {t}^{i} \left(t-1 \right)^{i} \,  P_i(S_0, \, t) 
 \cdot { S_1}^{i}
\,\, \, = \, \, 0
\end{eqnarray}
where the  $\, P_i(S_0, \, t)$'s are polynomials in $\, t$
 and $\, S_0\, =\,\sigma$, 
of degree $\, 2\, i$ in $\, S_0$.   
Again, these relations (\ref{ratioN}) seen as algebraic curves in
$\,S_0$ and $ \, S_1$ ($\, t$ being seen as a parameter), are
{\em rational curves}. From the previous remark that the $\, C(N, \, N)$ are
homogeneous polynomials of $\, E$ and $\, K$ one can easily deduce that 
 $\, S_0 \,= \, \sigma$ and  $\, S_1 \,= \, \sigma'$ are rational expressions
of the ratio $\,r \, = \,  E/K$ (or $\, E'/E$).

Now, similarly to the previous calculations, recalling
 that the variable $\, S_1 $ is 
 the derivative with respect to $\, t$ of the  variable $\, S_0$, 
 one also finds Riccati equations similar to (\ref{Ricatti})
for the uniformizing parameter $\, u$:
\begin{eqnarray}
\label{Riccatti2}
{\frac{d\,u}{dt}}\,
=\, \, \beta_2(t) \cdot {u}^{2} + \, \,\beta_1(t) \cdot u \,  + \beta_0(t) 
\end{eqnarray}
where  $\, \beta_0(t)$,  $\, \beta_1(t)$ and  $\, \beta_2(t)$ are
quite simple rational expressions of $\, t$, the  Riccati equation
(\ref{Riccatti2}) having only $\, t\, = \, 0$, $\, t\, = 1$
and $\, t\, = \, \infty$ as regular singularities.
The calculations are too large to be given here and will be
detailed in a forthcoming publication.

Note that, in such ``global'' Riccati algebraic approach, one has to be careful
because of the existence of many singular\footnote{We use here 
the terminology of singular solutions of differential
 equations \cite{Ritt,Ritt1}.} solutions of (\ref{jimbo-miwa}) 
corresponding to algebraic $\, \tau$ functions :
\begin{eqnarray}
&&\sigma \, = \,   t(t-1) \cdot {\frac{d}{dt}}\log(\tau)\,-1/4  \\
&&   \tau \, = \,  t^{\alpha} \cdot (1-t)^{\beta}, \qquad 
\left( 4\,\beta-1 \right)^{2}{N}^{2}+16\,\beta \left( 4\,\alpha+1 \right) 
 \left( \alpha+\beta \right) \, = \, \, 0 
\nonumber 
\end{eqnarray}
like, for instance, $\, (\alpha, \, \beta)$ being 
 $\, (-N/2 , \, -1/4\cdot N/(N-1))$,  
 $\, (-1/8\cdot (4\, N^2+1), \, N^2)$ or $\, (-1/4, \, 1/4)$, and especially
 $\, (N/2, \, -1/4\cdot N/(N+1))$
which corresponds to a series expansion with leading order similar
to (\ref{hyper}).

\section{Generalization to non-diagonal correlation functions $\, C(N, \, M)$}
\label{gener}

Most of the results, previously displayed, can be generalized to the
non-diagonal correlation functions $\, C(N, \, M)$ of the square Ising
model. The  $\, C(N, \, M)$'s are also given by determinants
(see \cite{mo-po-wa-63}) whose entries are holonomic quantities solutions
of linear differential equations of order three.
The $\, C(N, \, M)$'s are thus holonomic solutions of linear differential
equations. At first sight the growth of the order of the corresponding
differential operators should also be exponential in $\, N$ and  $\, M$.

We found that the order of these  linear differential operators
is, again, not growing exponentially with $\, N$ and $\, M \,$ but has a
{\em quadratic growth order} and depends on the parity of $\, M-N$.
For all the Fuchsian linear differential operators we have
obtained ($N$ and $\, M \, \le 6$), the order
can be reproduced  by :
\begin{eqnarray}
q\,=\,\, \,{{1} \over {8}}\cdot  \left( M+N+2 \right) \cdot 
 \Bigl( 4\, + \left( 3\, - \, (-1)^{M-N} \right)
\cdot   \vert M-N \vert  \Bigr) 
\end{eqnarray}
These linear differential operators $\, L_{NM}$ are too large to be given 
explicitly here. 
Let us just give one of them, namely the linear differential operator
$\, L_{12}$, corresponding  to the simplest non-diagonal (and non
horizontal or  vertical like $ C(0, \, N)$ or $ C(N, \, 0)$)
two-point correlation function.
The linear differential operator $\, L_{12}$ reads
\begin{eqnarray}
&& L_{12} \, = \, \,  D_s^{5}\, 
+{\frac { 5\left( 2\,s^2+3 \right)  D_s^{4}}{s \left( 1+s^2 \right) }}
\, +{\frac {q_3 \cdot  D_s^{3}}{ s^2 (1+s)^2 (1-s)^2 (1+s^2)^2}}\,
  \nonumber \\
&& \qquad +{\frac { q_2 \cdot D_s^{2}}{s^3 (1+s)^3 (1-s)^3 (1+s^2)^3 }}\, 
+{\frac { q_1 \cdot  D_s}{s^4 (1+s)^3 (1-s)^3 (1+s^2)^4}}\, \nonumber \\
&& \qquad  \quad +{\frac { q_0}{ s^5 (1+s)^3 (1-s)^3 (1+s^2)^5}}
\end{eqnarray}
where the polynomials $q_i$ read :
\begin{eqnarray}
q_3 &=& 13\,{s}^{8}+30\,{s}^{6}-78\,{s}^{4}-50\,{s}^{2}+53   \\
q_2 &=& 5\,{s}^{12}-7\,{s}^{10}+34\,{s}^{8}-128\,{s}^{6}
-65\,{s}^{4}-97\,{s}^{2}+2 \nonumber \\
q_1 &=& -5\,{s}^{14}+2\,{s}^{12}-67\,{s}^{10}
-118\,{s}^{8}-816\,{s}^{6}+157\,{s}^4-76\,s^2-101 \nonumber \\
q_0 &=& -192\,{s}^{10}+1840\,{s}^{8}-453\,{s}^{6}
+127\,{s}^{4}-15\,{s}^{2}-27 \nonumber 
\end{eqnarray}

Let us comment on the remarkable simplifications we encountered when
computing the $\, C(N, \, M)$'s from the quadratic double recursions
(discrete generalizations of Painlev\'e equations)
they satisfy~\cite{or-ni-gu-pe-01b} together with the $\, C^{*}(N, \, M)$'s .
From the expressions of the $\, C(N, \, N)$'s as  homogeneous polynomial in
 $\, E$ and $\, K$, and the  expressions of $\, C(0, \, 1)$,
 we can obtain
the  $\, C(N, \, M)$ and  $\, C^{*}(N, \, M)$, step by step using this
quadratic double recursion~\cite{or-ni-gu-pe-01b}.
At first sight these $\, C(N, \, M)$'s should be given as rational
expressions of $\, E$ and $\, K$
and,  in some cases, as roots of quadratic polynomials with polynomial
expressions in  $\, E$ and $\, K$.
Remarkably, as a consequence of factorizations and simplifications
in the numerator and denominator of these rational expressions, and
the occurrence of a perfect square in the
case of   roots of quadratic polynomials, the $\, C(N, \, M)$'s
are {\em actually} always given
by polynomial expressions in  $\, E$ and $\, K$, that are no longer
homogeneous polynomials,
 but {\em sums} of homogeneous polynomials\footnote[8]{This result can also be found 
in the Eqs. (3.22)-(3.35) of~\cite{Odyssey2}, which 
first show that
$C(N-1,N)$  is a homogeneous polynomial of E, K and the complete elliptic integral
of the third kind $\, \Pi_1$ in the anisotropic case. Together with
information from Montroll, Potts and Ward~\cite{mo-po-wa-63} (note e.g. eq. A19)
this means the same statement holds for $\, C(N-k,N)$ for $\, k=\, 2,...,N$.
Reduction of $\, \Pi_1$ in the isotropic case then shows that $\,C(N-k,N)$
is an inhomogeneous polynomial of E and K.}, as the following example
 shows\footnote[5]{Our results on the expressions of the $\, C(N, \, M)$'s
 are in agreement with those given , for $N$ and $M$ $\le 4$, in \cite{Ghosh2,Ghosh3}.}:
\begin{eqnarray}
\label{c13}
&&C(1, \, 3) \, = \, \, {{1} \over{3 \, s^6}}
 \cdot (P_1 \, + \, P_3)  \\
&& P_1\, = \, \,  2\, \left( {s}^{4}-1 \right) 
 \left( {s}^{2}+1 \right) {s}^{2} \cdot  K\, 
-{s}^{2} \left( {s}^{2}+1 \right)  \left( {s}^{4}
+3\,{s}^{2}-2 \right) \cdot  E   
         \nonumber \\
&& P_3\, = \, \,   \left( 6\,{s}^{2}-1+11\,{s}^{4} \right) \cdot  E^{3}
+ \left( {s}^{4}-1 \right)  
\left( 7\,{s}^{4}+12\,{s}^{2}-3 \right) \cdot  K\,E^{2}\, \nonumber \\
&& \quad  + \left( {s}^{4}-1 \right)  \left({s}^{2}+3 \right) 
 \left( {s}^{4}+2\,{s}^{2}-1 \right)  
\left( {s}^{2}-1 \right) \cdot  E\, K^{2}\,\nonumber \\
&&  \quad  + \left( {s}^{4}-1 \right)^{2} 
\left( {s}^{2}-1 \right)^{2} \cdot  K^{3}
\nonumber
\end{eqnarray}
The two linear and cubic components $P_1/3s^6$ and $P_3/3s^6$ are
respectively solutions of the two linear differential operators:
\begin{eqnarray}
&&L_1 \, = \, \,  D_s^{2}\, -{\frac { \left( 3\,{s}^{4}-7\,{s}^{2}+14 \right)
 }{s \left( {s}^{2} +1 \right)  \left( {s}^{2}-2 \right) }}\cdot D_s \, 
+4\,{\frac {11\,{s}^{4}-9\,{s}^{2}+4}{{s}^{2} \left( {s}^{2}+1 \right)^{2} 
\left( {s}^{2}-2 \right)  \left( -1+{s}^{2} \right) }}
 \nonumber \\
&&L_3 \, = \, \, D_s^4 \,
 - 2 \cdot  {{A_3} \over{(s^2-1)\, s  \cdot N}}  \cdot D_s^3\,
 + {{A_2 } \over { s^2 \, (s^4-1)^2   \cdot  N }} \cdot D_s^2  \nonumber \\
&&\quad \, 
+ {{A_1 } \over { s^3 \, (s^4-1)^2   \cdot  N }} \cdot D_s \, 
 + {{A_0 } \over { s^4 \, (s^4-1)^3   \cdot  N }}
\\
&&N \, = \, \, {s}^{12}+5\,{s}^{10}
+14\,{s}^{8}+54\,{s}^{6}+49\,{s}^{4}+13\,{s}^{2}-1
\nonumber \\
&&A_3  \, = \, \, 3\,{s}^{14}+15\,{s}^{12}+44\,{s}^{10}
+98\,{s}^{8}+383\,{s}^{6}+415\,{s}^{4}+133\,{s}^{2}-11
\nonumber\\
&&A_2  \, = \, \,19\,{s}^{20}+121\,{s}^{18}+248\,{s}^{16}
-408\,{s}^{14}-974\,{s}^{12}+2546\,{s}^{10} \nonumber \\
&&\quad 
+9597\,{s}^{8}+11440\,{s}^{6}+6521\,{s}^{4}+1277\,{s}^{2}-147  \\
&&A_1  \, = \, \, -27\,{s}^{20}-161\,{s}^{18}+240\,{s}^{16}
+5576\,{s}^{14}+17854\,{s}^{12}
+28590\,{s}^{10} \nonumber \\
&&\quad +30491\,{s}^{8}+19360\,{s}^{6}+8799\,{s}^{4}+1931\,{s}^{2}-333
\nonumber\\
&&A_0  \, = \, \,
-1792\,{s}^{20}-13136\,{s}^{18}-37568\,{s}^{16}-52256\,{s}^{14}
-48848\,{s}^{12} \nonumber \\
&&\quad -32576\,{s}^{10}-20720\,{s}^{8}-1568\,{s}^{6}+1600\,{s}^{4}
-688\,{s}^{2}+192
 \nonumber 
\end{eqnarray}
which are homomorphic to the first and third symmetric power of the
linear differential operator $\, L_E$:
\begin{eqnarray}
&&  L_{3} \, \, \, \,  {\rm equiv.}  \, \, \,  Sym^3(L_{E}), \qquad 
\hbox{that is :}  \qquad     L_{3} \cdot Q_3   \, = 
\, \,  W_3 \cdot  Sym^3(L_{E}) \nonumber \\
&&  L_{1} \, \, \, \,   {\rm equiv.} \, \, \,  L_{E}, \qquad 
\hbox{that is :}  \qquad   
  L_{1} \cdot Q_1   \, = \, \,  W_1 \cdot  L_{E}   
\end{eqnarray}
where $\, Q_3$ and $\, W_3$ (resp. $\, Q_1$ and $\, W_1$)
 are linear differential operators of order three (resp. one).
The order six linear differential operator corresponding to $\, C(1, \, 3)$,
that is the LCLM of  $\, L_1$ and  $\, L_3$ is 
homomorphic to the  LCLM of $\, L_E$ and 
$\,  Sym^3(L_{E})$ :
\begin{eqnarray}
L_1 \oplus L_3 \, \, \, \, \,\quad   
 {\rm equiv.} \, \, \, \quad   L_E \oplus  Sym^3(L_{E})
\end{eqnarray}

Also note that for the horizontal, or vertical,
 correlations ($\, N\, = 0$ or $\, M\, = 0$)
one also has a homogeneous polynomials of $\, E$ and 
$\, K$  of degree zero. Let us consider for instance
the simple correlation $\, C(0, \, 1)$ :
\begin{eqnarray}
\label{c01}
C(0, \, 1) \, = \, \, \, 1/2\,{\frac {\sqrt {1+{s}^{2}}}{s}} \, 
+ \, 1/2\,{\frac {\, \left( s-1 \right)  
\left( s+1 \right) \sqrt {1+{s}^{2}}}{s}} \cdot K
\end{eqnarray}
The first term (of degree zero in  $\, E$ and $\, K$) is solution of 
an order one linear differential operator $\, l_0$, whereas the second term is 
solution of an order two linear differential operator $\, l_1$:
\begin{eqnarray}
l_0 \, = \, \, \,{\it Ds} \, +{\frac {1}{s \left( 1+{s}^{2} \right) }}
,  \\
l_1 \, = \, \, \,{{\it Ds}}^{2}\, 
+{\frac { \left( {s}^{2}-3 \right) {\it Ds}}{s \left( {s}^{2}-1 \right) }}  \, 
+{\frac {2\,{s}^{6}+9\,{s}^{4}+4\,{s}^{2}+1}{
 \left( 1+{s}^{2} \right) ^{2}{s}^{2} \left( {s}^{2}-1 \right)^{2}}}.
\nonumber
\end{eqnarray}
Up to a conjugation by $\, (1+s^2)^{1/2}$, the 
order two linear differential operator $\, l_1$ is an operator 
homomorphic to $\, L_E$:
\begin{eqnarray}
&&(1+s^2)^{-1/2}   \cdot l_1 \cdot (1+s^2)^{1/2} \, \, = \, \,  \\
&&{{\it Ds}}^{2}\, 
+{\frac { \left( -4\,{s}^{2}+3\,{s}^{4}-3 \right)}
{ \left( 1+{s}^{2} \right) s \left( {s}^{2}-1 \right) }} \cdot  {\it Ds}
\, 
+{\frac {{s}^{6}-{s}^{4}+7\,{s}^{2}+1}{ \left( {s}^{2}-1 \right)^{2} 
\left( 1+{s}^{2} \right) {s}^{2}}} \quad 
 \nonumber \quad 
\end{eqnarray}
with $\, (1+s^2)^{-1/2}   \cdot l_1 \cdot (1+s^2)^{1/2} \, \,  {\rm equiv.} \, \, \,\,   L_E$.
One actually finds that $\, C(0, \, 1)$ is solution of the third 
order operator direct sum of $\,l_0 $
and   $\,l_1$ and is thus equivalent (up to  conjugation by $\, (1+s^2)^{1/2}$)
 to the  direct sum of $\,l_0 $ and $\, L_E$.

\vskip 0.2cm

From the fact that the $\, C(N, \, M)$'s are {\em actually} always given 
by polynomial expressions  sums of homogeneous polynomials in
$\, E$ and $\, K$, one
easily deduces that the corresponding linear differential operators
$\, L_{NM}$
are homomorphic to direct sums of symmetric products of the
 second order linear differential operator (\ref{Lee}),
 yielding generalizations of (\ref{Sym}):
\begin{eqnarray}
\label{SymP}
L_{NM} \, \, \, \, \quad  {\rm equiv.} \, \, \quad \, \oplus_m Sym^m(L_{E})
\end{eqnarray}
where for $N-M$ odd, $m$ is running as $N, N+1, N+2, \cdots ,M$ and for
$N-M$ even, as $N, N+2, N+4, \cdots ,M$, and where
 $ \, Sym^m(L_{E})\, = \, l_ 0\, $ when $\, m\,=\, 0$.

This structure is a consequence of the fact that the $\, C(N, \, M)$'s
are  given
by {\em polynomial expressions} in  $\, E$ and $\, K$,
 instead of the {\em rational or
algebraic
expressions} in  $\, E$ and $\, K$ which one could expect at first sight from
the discrete Painlev\'e double recursions.
This corresponds to quite remarkable identities and simplifications
(factorizations, occurrence of perfect squares).
From a less non-linear and more  ``Fuchsian'' linear viewpoint, an explanation
is the following. The  non-diagonal $\, C(N, \, M)$
 are determinants of holonomic
functions, hence they are holonomic themselves. On the other hand, they 
are rational (or even algebraic expressions in $E$ and $K$). Now, because 
the Galois group of $L_E$ is $SL(2,C)$, 
results from \cite{PutSinger,Singer2} show that expressions in
 $E$ and $K$ which are holonomic will have to
be polynomial.

Again one can check that all these linear differential operators
$\, L_{NM}$  are Fuchsian differential operators with only three regular
singular points $\, t \, =\, 0$,  $\, t \, =\, 1$,  $\, t \, =\, \infty$.
This is a straight consequence of the fact that these $\, L_{NM}$'s can be
built as linear differential operators having polynomial solutions in
$\, E$ and $\, K$ and thus, 
they inherited the three regular singular points $\, t \, =\, 0$,
  $\, t \, =\, 1$,  $\, t \, =\, \infty$
from the complete elliptic integrals $\, E$ and $\, K$, and from the fact
that the coefficients of the
monomials $\, E^i \cdot K^j$ are extremely simple rational expressions with no
singularity except poles at $\, s=0$
(polynomial in $\, s$ divided by powers of $\, s$).

The results we got on the non-diagonal correlation functions $\, C(N, \, M)$
are too numerous, and
require too much space, to be given here (even if the final result is
remarkably simple and elegant). 
However one sees the emergence of quite fascinating structures relating 
an infinite set of Fuchsian linear differential operators depending on two
integers $\, N$ and $\, M$ (the $\, L_{NM}$'s),
with some quadratic double recursions that are nothing but discrete
generalizations of Painlev\'e, these structures
being themselves closely linked with complete elliptic integrals.

\section{B\"acklund transformation and Malmquist Hamiltonian structure}
\label{backlund}

Let us recall that since the work of Malmquist~\cite{Malmquist} it has
been known that Painlev\'e VI equation can be obtained from Hamilton
equations
\begin{eqnarray}
p\, \prime ={{d\, p} \over {d\,t}} \, = \, \, 
-{{\partial H} \over {\partial q}} , \, \, \quad  
q\, \prime={{d\, q} \over {d\,t}} \, = \, \,
 {{\partial H} \over {\partial p}}
\end{eqnarray}
with
\begin{eqnarray}
&&t\, (t-1) \cdot H \,  
 \, = \, \,q \left( q-1 \right)  \left( q-t \right) {p}^{2}\, -\, Q( q)\cdot p \,   \\
&& \qquad  + \left(  n_3- n_1 \right) 
 \left(  n_3- n_2 \right)  \left( q-t \right), 
\qquad \hbox{where :} \qquad Q( q)\, = \, \, \nonumber \\
&&  \left( n_3
+ n_4 \right)  \left( q-1 \right)  \left( q-t \right) + \left(
 n_3- n_4 \right) q \left( q-t \right)
 - \left(  n_1+n_2 \right)  \left( q-1 \right) q 
 \nonumber
\end{eqnarray}
With this structure, it follows that $p$ is a rational function of $t$, $q$
and $q\,\prime$. The Hamiltonian is the $t-$logarithmic derivative of
the function $\tau(t)$. The correlation functions $C(N,N)$ being solutions
of the sigma form of Painlev\'e VI, one may find how
the expressions of the two variables $\,p$ and $\,q$ (for which
the B\"acklund transformations are {\em birational})
in the restricted case  $n_1\, =\, N/2$, $ n_2\, =\, (1-N)/2$, 
$n_3\, =(1+N)/2$ and $n_4\, =N/2$ appear in terms of the elliptic
integrals $K$ and $E$.
Considering the diagonal correlation function $C(2, \, 2)$ taken
as $\tau(t)=t^{1/4}\,C(2,\,2)$
one might expect, at first sight, to obtain the  variables
$\,p$ and $\,q$ as algebraic expressions in terms of 
$\, E$ and $\, K$ (and $ \, t$).
Remarkably, one obtains the surprising result that the two variables
$\,p$ and $\,q$ are {\em actually rational expressions\footnote[5]{Formulas 
expressing $\, p$ and $\, q$ as
ratios of tau functions can be found in  eqs. (5.42), (5.43)
of Forrester and Witte~\cite{Forrest2}.
} of } $\, E$ and $\, K$.
For $\, N=2$ one thus gets two solutions, the simplest one being:
\begin{eqnarray}
&&p = -{\frac{\left( \left( t+1 \right) E
+ \left( t-1 \right) K \right) \, N_p^{(1)} \cdot N_p^{(2)} }
{2t\,\left( 2\,E+ \left( t-1 \right) K \right)
 \,D_p^{(1)} \,D_p^{(2)}}}, \quad \nonumber \\
&&q = -{\frac{ t\, \left( 2\,E+ \left( t-1 \right) K \right)  \cdot N_q }
{\left(  \left( t+1 \right) E+ \left( t-1 \right) K \right) \, N_p^{(1)}}}
\end{eqnarray}
\begin{eqnarray}
&&N_p^{(1)} =
- \left( 9\,t-1 \right)  \left( t-1 \right)^{2}\cdot {K}^{2}-2\, \left( 17
\,t-1 \right)  \left( t-1 \right) \cdot EK\nonumber     \\
&&  \qquad + \left( 1+{t}^{2}-34\,t \right)\cdot {E}^{2}  \nonumber     \\
&&N_p^{(2)} =
  - \left( t-1 \right) {K}^{2}-2\,EK+{E}^{2}  \\
&&D_p^{(1)} =
    -3\,{K}^{2} \left( 
t-1 \right) ^{2}-8\, \left( t-1 \right) EK+ \left( -5+t \right) {E}^{2
} \nonumber \\
&&D_p^{(2)} =
-{K}^{2} \left( t-1 \right) ^{2}+2\, \left( t-1 \right) ^{2}EK+
 \left( 5\,t-1 \right) {E}^{2} \nonumber \\
&&N_q =
- \left( 3\,t-11 \right)  \left( t-1 \right)^{2} \cdot {K}^{2}\, 
+2\, \left( t-1 \right)  \left( 3\,{t}^{2}-t+14 \right)\cdot  EK \nonumber \\
&& \qquad + \left( 17\,{t}^{2}-2\,t+ 17 \right)\cdot  {E}^{2} \nonumber
\end{eqnarray}

One notes the homogeneous occurrence, in terms of degree, of $E$ and $K$
in these relations.
The variables $p$ and $q$ have the rational parametrization of an algebraic
curve. Obviously, the uniformization parameter similar to the one introduced
in Sec. \ref{conditions} can be chosen as the  ratio
$ u \, = \, \, E/K$ (or $\, E'/E$) of the two elliptic
 integrals. One can then deduce that
the parameter $u$ is a solution of a Riccati differential equation.
These results generalize straightforwardly to all the $\, p$, $\, q$  
associated with the  $\,C(N, \, N)$'s leading, 
remarkably, to  {\em rational functions} of $\, E$ and $\, K$ and
yielding rational parametrization
for the corresponding algebraic curves between $\, p$ and $\, q$. 
We have the same results in the variables $\, \sigma $ and $\, \sigma' $.
The expressions of the B\"acklund transformation corresponding to 
changing $N$ into $N+1$ in terms of the variables $\, p$, $\, q$
will be analyzed elsewhere.

\section{Conclusion}
\label{concl} 
The phenomenon of the existence of a one parameter family of solutions
to Painlev{\'e} VI equation  has been presented in this paper by the
study of the specific PVI equation which is
  satisfied by $C(N,N)$ the diagonal correlation function of the Ising
  model. However the existence of such linear equations is a much
  larger phenomena and certainly holds for all PVI equations where
the difference of any two of the parameters $v_j$ is an integer
because, in that case, there is a class of solutions which can be
written as finite dimensional determinants whose elements are
hypergeometric functions.

Even though the existence of these Fuchsian
differential equations follows from the general theorem on holonomic
functions the specific form and properties of these
equations is tedious to obtain. 
However, the expressions obtained for small $N$ 
(via series computations) have been sufficient to guess the 
structure that is proved in sections 2 and 3.
Moreover, using these initial computations, it has been possible to
make a remarkably simple conjecture for the exponents which is in
complete agreement with the local expansion of the Painlev{\'e} VI
equation at its singular points and this conjecture puts restrictions
on the coefficients in the differential equations. 

In this paper we have obtained the Fuchsian equations by starting
with the PVI equation. However the question can be reversed and we can
ask what are the conditions on the Fuchsian equations which will lead
to PVI equations. For second order Fuchsian equations it would be
sufficient to require that the exponents at the singularities agree
with the exponents allowed by the local expansions of PVI. 
But for higher order
equations the exponents do not fully specify the Fuchsian equation. 
The extra parameters which need to be
specified are referred to as accessory parameters and
only very specific accessory parameters will lead to Fuchsian
solutions of PVI. The needed restrictions on these parameters
are not known. 

The  more general version of this  is the question of 
determining whether or not a specific set of solutions to a Fuchsian equation
will also satisfy some nonlinear equation (not necessarily PVI).
This is in some sense the original
question asked by Jimbo and Miwa \cite{jim-miw-80}
and this is particularly important because, for  $C(N,N)$, the nonlinear PVI
is much simpler than the linear equations $L_{N,N}$.
It was found in~\cite{ze-bo-ha-ma-04,ze-bo-ha-ma-05,ze-bo-ha-ma-05b,ze-bo-ha-ma-05c} that the
three and four particle contributions to the susceptibility 
of the Ising model, $\chi$,
satisfy Fuchsian equations whose structure appears rather complicated
and the question may be asked whether these functions, 
or their sum $\chi$, can also satisfy
a nonlinear equation 
which might be simpler in appearance.

Finally we remark that perhaps the most interesting 
discovery in this paper is that the
operator $L_{NN}$ are equivalent to the $N^{th}$ symmetric power of the
operator $L_E$. 
This property extends to the operator $L_h$ (which is isomorphic to $L_E$).
One might wonder whether all solutions of the sigma form of Painlev\'e VI that are 
also solutions of linear differential equations would be produced from 
symmetric powers of $L_E$ by intertwinners.

\end{document}